\documentclass[english]{emulateapj}
\usepackage{lmodern}

\usepackage[T1]{fontenc}
\usepackage[latin9]{inputenc}
\usepackage{color}
\usepackage{array}
\usepackage{float}
\usepackage{multirow}
\usepackage{wasysym}
\usepackage{graphicx}
\usepackage{esint}

\makeatletter

\providecommand{\tabularnewline}{\\}

\makeatother

\usepackage{babel}
\begin{document}

\title{The History of Tidal Disruption Events in Galactic Nuclei}

\author{Danor Aharon, Alessandra Mastrobuono Battisti \& Hagai B. Perets}

\affil{Physics Department, Technion - Israel Institute of Technology, Haifa,
Israel 3200003}
\begin{abstract}
The tidal disruption of a star by a massive black hole (MBH) is thought
to produce a transient luminous event. Such tidal disruption events
(TDEs) may play an important role in the detection and characterization
of MBHs and probe the properties and dynamics of their nuclear stellar
clusters (NSCs) hosts. Previous studies estimated the recent rates
of TDEs in the local universe. However, the long-term evolution of
the rates throughout the history of the universe has been hardly explored.
Here we consider the TDE history, using evolutionary models for the
evolution of galactic nuclei. We use a 1D Fokker-Planck approach to
explore the evolution of MBH-hosting NSCs, and obtain the disruption
rates of stars during their evolution. We complement these with an
analysis of TDEs history based on N-body simulation data, and find
them to be comparable. We consider NSCs that are built-up from close-in
star-formation (SF) or from SF/clusters-dispersal far-out, a few pc
from the MBH. We also explore cases where primordial NSCs exist and
later evolve through additional star-formation/cluster-dispersal processes.
We study the dependence of the TDE history on the type of galaxy,
as well as the dependence on the MBH mass. These provide several scenarios,
with a continuous increase of the TDE rates over time for cases of
far-out SF and a more complex behavior for the close-in SF cases.
Finally, we integrate the TDE histories of the various scenarios to
provide a total TDE history of the universe, which can be potentially
probed with future large surveys (e.g. LSST). 
\end{abstract}

\section{INTRODUCTION}

A tidal disruption event (TDE) by a massive black hole (MBH) occurs
when a star comes closer than its tidal radius (approximately given
by $r_{t}=(M_{\bullet}/m_{\star})^{1/3}r_{\star};$ where $M_{\bullet}$
is the mass of the MBH, $m_{\star}$ is the mass of the star and $r_{\star}$
is its radius). The star is then pulled apart by the tidal forces
by the MBH, part of its material is ejected away and part is accreted
on the MBH \citep[e.g.][]{1988Natur.333..523R}. The explosive disruption
itself, and the later accretion on the MBH likely give rise to transient
energetic phenomena potentially observable (with more than ten potential
detections to date) with current and future telescopes \citep[e.g. ][]{2014PhT....67e..37G}. 

MBHs are thought to exist in a significant fraction of all galactic
nuclei. Due to their high mass ($10^{5}-10^{9}M_{\odot}$), they dominate
the gravitational potential in the nuclear star clusters (NSCs) which
typically host them (at least at the $10^{5}-10^{8}$ $M_{\odot}$
MBH mass regime); the orbits of stars in the central parsecs around
MBHs are therefore governed by the potential of the MBHs. The long-term
dynamical evolution of the stars in the region of influence of the
MBH is also determined by their interactions and mutual gravitational
scatterings with other stars in the NSC. These can be modeled using
a Fokker-Planck (FP) approach to describe the diffusive behavior of
stars in NSC. (\citealt{1976ApJ...209..214B}), where the close approaches
of stars to the MBH leading to their tidal disruption can be modeled
as a sink-term in the FP equations. TDE rates calculated following
such an approach typically give rise to rates of the order of $10^{-5}$
up to $10^{-3}yr^{-1}$ per galaxy in the local Universe, assuming
some steady state condition of the NSC (e.g. \citealp{Wang2004},
and references therein).

Since the 1980's tidal disruption of stars by MBHs were suggested
to occur in galactic nuclei and produce luminous transient events
with unique signatures \citep{1988Natur.333..523R}. The signatures
can teach us both on the dynamics in galactic nuclei as well as on
the conditions very close to MBHs, and their properties. Such TDEs
have since been extensively studied theoretically, and a significant
observational effort has been put in their detection \citep{2014PhT....67e..37G}.
Many studies have estimated the rates of TDEs (e.g. \citealp{1999MNRAS.309..447M},
\citet{Merritt2009} and \citealt{2011MNRAS.418.1308B}) in the local
universe, and several estimates have been derived based on direct
observations of TDE-candidates in recent years (\citealt{2014arXiv1407.6284K,2014arXiv1407.6425V}).
Since the structure and the properties of NSCs vary with time, it
is expected that the TDE rates will change accordingly. Moreover,
the change of TDE rates can be used as a proxy for characterizing
the properties of NSCs and/or the history of the build-up of MBHs
and NSCs in the universe. In particular, future deep surveys may allow
the detection of TDEs from earlier stages of cosmological evolution
arising from younger galaxies, and potentially provide a handle on
the TDE history \citep{2014PhT....67e..37G}. Nevertheless, the evolution
of TDE rates with time and their dependence on the NSC and star-formation
history and evolution have been little studied. 

In this work, we explore the TDEs history (TDH) by studying the rate
of tidal disruption of stars in NSCs. We simulate the evolution of
NSCs in several types of galaxies by FP analysis as well as N-body
simulations. We obtain the TDE rates as a function of their evolutionary
history, and find that they are affected by the NSC structure and
its build-up history. By integrating the TDE rates from galaxies of
each type we obtain the total galaxy averaged TDH of the universe
expected to be observed. We thereby provide a first global estimate
of the TDH accounting for NSC build-up, albeit with simple models. 

In the following we begin by a brief description of the Fokker-Planck
analysis and N-body simulations of evolving clusters. We describe
our approach of evaluating a sink term from FP analysis for calculating
the tidal disruption rate. We then present the TDH and its dependence
on the MBH mass and the NSC build-up history in different types of
galaxies. Finally, we consider the global, galaxy averaged, TDH of
the universe evaluated by integrating the TDE rates arising from the
different types of galaxies (and their correspondent evolutionary
histories) and relevant MBH masses. We note, that our model is based
on typical NSCs that undergo basic processes and their dynamics are
dominated by two-body relaxation. The model can be expanded, and future
works may include other processes and properties in NSC such as non
uniform stellar mass, self-consistent potential, non-spherical NSCs
etc. Here we provide a first study of the history of TDEs using simple
models.

\section{METHODS}

In our work we explore the tidal disruption rates in NSCs through
their evolution in time. We focus on simple, spherical NSCs that evolve
through in-situ star formation (SF) \citep{2015ApJ...799..185A},
and study other cases of NSCs that are built-up through consecutive
infalls of massive clusters or evolve from a pre-existing stellar
cusp. The main analysis method used in our work is based on the numerical
solution of the FP equation, complemented by N-body simulations of
NSCs evolution.

Our model includes a stellar cluster, the NSC, harboring a central
MBH. It focuses on the stars in the central few parsecs of the NSC,
and in particular in the range between the tidal radius - $r_{t}\approx r_{*}(M_{\bullet}/M_{*})^{1/3}$
below which regular stars are disrupted by the MBH, and the radius
of influence where the stellar motions are dominated by the MBH potential,
defined by 

\begin{equation}
r_{h}=GM_{\bullet}/\sigma_{\star}^{2},\label{eq:r_h}
\end{equation}
 where $r_{*}$, and $M_{*}$ are the typical radius and mass of stars
in the NSC, respectively; and $\sigma_{\star}$ is the velocity dispersion
of stars just outside the NSC. For $r>r_{h}$ the original model following
\citet[hereafter BW]{1976ApJ...209..214B} assumes the existence of
a ``thermal bath'' which supplies stars to the inner region of the
galactic nucleus. The stellar orbits within $r_{h}$ are
assumed to be Keplerian in this range. The relaxation time, that dominates
the timescale of the stellar cluster, is defined as
\begin{equation}
T_{r}\approx\frac{\sigma^{3}}{G^{2}M_{*}\rho\ln(\Lambda)}\label{eq:relaxation}
\end{equation}
Where $\rho$ is the stellar density, and $\ln(\Lambda)$ is the Coulomb
logarithm, a factor which is related to the scale of the system ($\ln(\Lambda)\approx10)$.

\subsection{Fokker-Planck analysis and N-body simulation }

The FP model, extensively used in our work, consists of a time and
and energy-dependent, angular momentum-averaged particle conservation
equation. It has the form:

\begin{equation}
\frac{\partial f(E,t)}{\partial t}=-AE^{-\frac{5}{2}}\frac{\partial F}{\partial E}-F_{LC}(E,T)+F_{SF}(E,T)\label{eq:Fokker_Planck}
\end{equation}

where 
\begin{equation}
A=\frac{32\pi^{2}}{3}G^{2}M_{*}^{2}\ln(\Lambda)
\end{equation}

The term $F=F[f(E),E]$ is related to the stellar flow, and plays
an important role in the evolution of the stellar cluster. It presents
the flow of stars in energy space due to two-body relaxation, it is
defined by:

\begin{equation}
F=\int dE'\left(f(E,t)\frac{\partial f(E',t)}{\partial E'}-f(E',t)\frac{\partial f(E,t)}{\partial E}\right)\left(\max(E,E')\right)^{-\frac{3}{2}}\label{eq:flow_eq}
\end{equation}

The addition of the source term that represents the SF has the form:

\begin{equation}
F_{SF}(E,T)=\frac{\partial}{\partial t}\left(\Pi(E)E_{0}E^{\alpha}\right)\label{eq:SF_sourceT}
\end{equation}

$\Pi(E)$ is a rectangular function, which boundaries correspond to
the region where new stars are assumed to from; $E_{0}$ is the source
term amplitude; and $F_{SF}$ is a power-law function with a slope
$\alpha$, defining the SF distribution in phase space. For a detailed
description of the Fokker-Planck analysis see \citet{2015ApJ...799..185A}
where the same analysis was used for studying the evolution of NSCs
with in-situ SF. Note that the sink term $F_{LC}(E,T)$ in Eq. \ref{eq:Fokker_Planck}
is used for evaluating the tidal disruption rates in our models. This
term represents stars with energies in the interval $(E,E+dE)$ that
flow into the MBH (mostly due to angular momentum change; the sink-term
represent the effective loss in energy space; see \citealp{1977ApJ...216..883B,1977ApJ...211..244L}),
and are therefore lost from the system (see \citealt{2013CQGra..30x4005M}
for detailed overview of loss cone analysis).

For modeling tidal disruption through N-body simulations, we follow
the same methods and similar assumptions, as well as make the same
use of the same code in \citet{2014ApJ...784L..44P} and \citet{2014ApJ...796...40M},
where the TDE rates are evaluated. For a detailed method explanation
along with the initial conditions of the infalling clusters and TDEs
estimation, see \citet{2014ApJ...796...40M}. In brief, we run direct
N-body simulations (using the $\phi$GRAPE code, \citealt{Harfst2007})
of the consecutive infall and merging of a set of 12 single-mass globular
clusters each with mass of $10^{6}$ M$_{\odot}$, inspiralling from
a galactocentric distance of 20pc.

\section{MODELS}

\subsection{Formation and evolution of NSCs}

In our study, we explore the TDH through two formation/evolutionary
models of NSCs: in-situ star formation and cluster infall. The study
is done by using the FP analysis to simulate an evolving NSC with
an addition of extra source term (see \citealt{2015ApJ...799..185A}
for the full analysis). In the in-situ SF model we consider two possible
cases: NSCs that are entirely built-up from the in-situ SF, and NSCs
that evolve from an initial steady state BW distribution (hereinafter
``built-up SF'' origin and ``primordial cusp origin'', respectively).
In both of the scenarios, the initial density profile is determined
from an arbitrary initial distribution function (DF). For the bound
stars inside the radius of influence in the primordial cusp scenario
the DF takes the form of $f(E,t_{0})\propto E^{0.25}$ (corresponding
to the BW steady-state cusp which has the form of $n\propto r^{-7/4}$).
The normalization of the number density at the radius of influence
for each scenario is given in table \ref{table_dens_MBH_mass}. We
consider different masses for the hosted MBH in their centers, where
the density profiles of the NSCs are normalized in the same way as
done in \citeauthor{Wang2004} (2004), where the parameters were taken
from typical galaxies (e.g. NGC 4551/4621). The velocity dispersion
(\textcolor{black}{which is taken into account in the calculation
of the radius of influence)} is evaluated from the M-Sigma relation:
$M_{\bullet}\propto\sigma^{4.36}$, \citep{2001ApJ...547..140M}. 

\begin{table}[H]
\begin{tabular}{|c|c|c|c|c|}
\hline 
MBH mass {[}$M_{\odot}${]} & $5\times10^{5}$ & $1\times10^{6}$ & $5\times10^{6}$ & $1\times10^{7}$\tabularnewline
\hline 
\hline 
density profile at $r_{h}$ \textbf{{[}${\rm pc^{-3}}$}{]}  & $6\times10^{3}$ & $1\times10^{4}$ & $4\times10^{4}$ & $1\times10^{5}$\tabularnewline
\hline 
\end{tabular}\label{table_dens_MBH_mass}

\caption{Density profile at $r_{h}$ of NSCs for different MBH masses}
\end{table}

In the built-up SF scenario the initial DF is evaluated according
to the term $F_{SF}$ (Eq. \ref{eq:SF_sourceT}), where the initial
number density is equal to the added number density every time step.
For both of the scenarios, The chosen slope of the SF function is
motivated by the observed power-law (\citealt{2009ApJ...703.1323D,2009ApJ...697.1741B})
distribution of young stars observed in the young stellar disk in
the GC. The DF of the unbound stars ($-\infty<r<r_{h})$ has the form
of a Maxwellian distribution. We study the TDE rates throughout the
evolution of NSCs and For each scenario we tested several types of
SF history corresponding to the galaxy type studied (see Section \ref{sub:TDEs-in-different}).
Though we use the FP approach for our modeling of the TDH, we also
complement our analysis with N-body simulations of the cluster-infall
scenario. As discussed later on, we find that models of in-situ SF
in the outer regions of an NSC evolve very similarly to those of cluster
infall, since the clusters shed their material in the same regions,
effectively providing a source term similar to that arising from SF.
Therefore our models of in-situ SF in the outer regions of an NSC
effectively well capture models of cluster-infall, at least in terms
of the TDE rates under the assumptions made in our study.

\subsection{Star formation history in different type of galaxies\label{sub:TDEs-in-different} }

We followed the evolution of several types of galaxies with different
SF scenarios. For each type of galaxy and SF scenario we evaluated
the TDE rates and their evolution. We considered the following simplified
cases: (1) An elliptical galaxy with SF occurring only at the early
stages of the evolution of the NSC; and (2) spiral galaxy with either
continuous SF or repeating bursts of SF occurring throughout the evolution
of the NSC to present days. The correspondence between galaxy type
and its typical SF history follows \citet{2014ARA&A..52..415M}. We
used a simplified model to describe the non-continuous SF star-burst
scenario, where we assume the bursts occur at equally spaced time
intervals (Gyr), at ten times the rate of the continuous SF models,
but only for 100 Myrs (as to have the same total SF averaged rate).
We summarize the types of galaxies and the SF scenarios simulated
in our work in Table \ref{tab:galaxies_prop}. Note, that we also
study different scenarios for the regions where SF may occur; either
in the range between $2.0-3.5{\rm pc}$ or in the range $0.05-0.1{\rm pc}$
from the MBH. The ranges that were chosen for SF in our work are motivated
by the observed existence of a possibly star-forming circumnuclear
gaseous disk in the range of 2-5pc in the Galactic center (similar
young stellar populations are observed in other galactic nuclei, e.g.
\citealp{2006AJ....132.2539S}). Most of the presented SF scenarios
are based on these observations. The close-in SF region ($<0.1{\rm pc}$)
is motivated by the existence of a nuclear stellar disk of young stars
very close to the MBH in the Galactic center (e.g. \citealt{2009ApJ...703.1323D,2009ApJ...697.1741B}).
The averaged rates of SF were set to $10^{-4}-5\times10^{-3}{\rm yr^{-1}}$,
motivated by the observed number of stars in the central parsecs of
galactic nuclei, as to obtain a total mass of $\sim10^{6}-5\times10^{7}$M$_{\odot}$
comparable to the masses of NSCs derived from observations of the
GC and extragalactic NSCs (\citealp{2006AJ....132.2539S}). 

In addition, we note that the range of of $2.0-3.5{\rm pc}$ is also
representative of the range where a disrupted cluster dispenses its
stars close to the MBH, thereby building-up the NSC stellar population
(e.g. \citealt{2012ApJ...750..111A,2014ApJ...784L..44P}; the exact
cluster tidal radius depends on the MBH and NSC mass and could have
a slightly larger range). Finally, we also considered different rates
of SF in the NSC to access their overall effects on the TDH. All models
used are summarized in Table 1. Note that bursting SF models where
SF occurs in the outer regions also serve as good proxies for the
cluster infall model, as we discuss later on.

\begin{table*}
\centering
\begin{tabular}{|c|c|>{\centering}p{2cm}|>{\centering}p{2cm}|>{\centering}p{2.5cm}|}
\hline 
{\footnotesize{}Model (Galaxy type)} & {\footnotesize{}SF rate ($yr^{-1})$} & Spatial regions of SF (pc) & {\footnotesize{}Epoch of SF} & {\footnotesize{}NSC Origin}\tabularnewline
\hline 
\hline 
\multirow{4}{*}{{\footnotesize{}$E$ (elliptical)}} & \multirow{2}{*}{{\footnotesize{}$5\times10^{-3}$}} & \multirow{2}{2cm}{$2.0-3.5$; $0.05-0.1$} & {\footnotesize{}0-1 Gyr} & {\footnotesize{}Primordial BW cusp}\tabularnewline
\cline{4-5} 
 &  &  & {\footnotesize{}0-1 Gyr} & {\footnotesize{}build-up }\tabularnewline
\cline{2-5} 
 & \multirow{2}{*}{{\footnotesize{}$10^{-3}$}} & \multirow{2}{2cm}{$2.0-3.5$} & {\footnotesize{}0-1 Gyr} & {\footnotesize{}Primordial BW cusp}\tabularnewline
\cline{4-5} 
 &  &  & {\footnotesize{}0-1 Gyr} & {\footnotesize{}build-up }\tabularnewline
\hline 
\multirow{2}{*}{{\footnotesize{}$S_{0}$ (spiral with continuous SF)}} & \multirow{2}{*}{{\footnotesize{}$5\times10^{-4}$}} & \multirow{2}{2cm}{$2.0-3.5$; $0.05-0.1$} & {\footnotesize{}0-10 Gyr} & {\footnotesize{}Primordial BW cusp}\tabularnewline
\cline{4-5} 
 &  &  & {\footnotesize{}0-10 Gyr} & {\footnotesize{}build-up }\tabularnewline
\hline 
\multirow{4}{*}{{\footnotesize{}$S_{B}$ (spiral with bursts of SF)}} & \multirow{2}{*}{{\footnotesize{}Bursts of $5\times10^{-3}$}} & \multirow{2}{2cm}{$2.0-3.5$; $0.05-0.1$} & \multirow{4}{2cm}{{\footnotesize{}bursts of 100 Myr every 1 Gyr}} & {\footnotesize{}Primordial BW cusp}\tabularnewline
\cline{5-5} 
 &  &  &  & {\footnotesize{}build-up }\tabularnewline
\cline{2-3} \cline{5-5} 
 & \multirow{2}{*}{{\footnotesize{}Bursts of $10^{-3}$}} & \multirow{2}{2cm}{$2.0-3.5$} &  & {\footnotesize{}Primordial BW cusp}\tabularnewline
\cline{5-5} 
 &  &  &  & {\footnotesize{}build-up }\tabularnewline
\hline 
\end{tabular}

\caption{\label{tab:galaxies_prop}Models for nuclear stellar clusters and
star formation in different types of galaxies.}
\end{table*}

We simulated the evolution of each NSC (described in Table \ref{tab:galaxies_prop})
with its corresponding TDE for four different masses of MBHs: $5\times10^{5}M_{\odot}$,
$1\times10^{6}M_{\odot}$, $5\times10^{6}M_{\odot}$ and $1\times10^{7}M_{\odot}$
in order to evaluate a general form of TDEs dependency on MBH mass.
For each studied mass of MBH the appropriate relaxation time corresponding
to the velocity dispersion (evaluated from M-Sigma relation, \citealt{2001ApJ...547..140M})
was determined, and used to produce the primordial clusters considered
in the relevant models. The rates of tidal disruptions were then evaluated
throughout the NSC evolution. The evolution of NSCs hosting MBHs of
higher masses is not discussed in this work, since the tidal radius
for MS stars falls below the Schwarzschild radius ($2GM_{\bullet}/c^{2}$).
Therefore, MBH with higher masses would not contribute to the TDE
rates of typical main-sequence stars. Lower mass MBHs may also exist,
though there is little observational evidence for their current existence,
but this is likely due to current observational limitations. Nevertheless,
such low mass MBHs may give rise to TDEs with significantly different
observational signatures, and we therefore limit our study only to
more massive MBHs.

\subsection{Assumptions and Limitations\label{sub:Assumptions-and-Limitations}}

The disruption rates of stars by MBHs depend both on the NSC properties
as well as on several possible physical processes. These processes
include for example the presence of massive perturbers, that refill
the loss cone (e.g. \citealt{2007ApJ...656..709P}), the evolution
of central MBHs binaries and the effects of non spherically symmetric
potentials in the nucleus (see \citealt{2013degn.book.....M} for
a comprehensive review). Some of these processes are efficient in
increasing the TDE rates, but typically operate only on short timescales
(e.g. binary MBH mergers), while the level of non-sphericity of galactic
nuclei is not well known. Our model focuses the TDEs for number of
SF scenarios including different rates and different ranges. A similar
TDE history model was suggested by \citet{Merritt2009} whose work
was based on different evolution scenarios of NSCs which does not
include SF. A comparison between Merritt's work and this study results
is discussed in section \ref{sub:Cluster-infall-and SF}.

Another important assumption is that the MBHs in our models evolve
and change their mass. This change is done by adding the mass of the
disrupted stars to the MBH during the evolution of an NSC. We emphasize,
however, that the growth of the MBH mainly affects the TDE rates in
the lower rates of SF scenarios and for initial low masses of MBH,
in which cases the total added mass becomes comparable to the initial
mass of the MBH . In these cases, the TDE rates decline with time
(see Fig. \ref{fig:Evolution-of-TDE_lower-rates-SF}), while the TDE
rates grow with time for the more massive MBHs.

We continuously change the NSC structure up to the radius of influence
according to the evolution of the FP models. As the mass of the MBH
grows we change the velocity dispersion $\sigma_{\star}$ assuming
the M-$\sigma$ relation holds at any given time. The radius of influence
(Eq. \ref{eq:r_h}) is then changed according to the updated mass
and velocity dispersion, and the outer boundary condition of the thermal
bath also change accordingly, following \citet{Wang2004}. We note
and caution that this is not a trivial assumption; its advantage is
in consistently relating the MBH, the velocity dispersion and the
stellar density outside the NSC, and therefore, by definition producing
a final NSC and environmental configuration which is in accordance
with the observed ${\rm M-\sigma}$ relation today (though the relation
is defined in the context of the velocity dispersion at larger radii).
The disadvantage is in using an ad-hoc assumption in which the ${\rm M-\sigma}$relation
is continuously kept at all times; whether this holds is currently
unknown. To the best of our knowledge previous studies used either
a constant outer boundary conditions \citep[e.g. ][]{1976ApJ...209..214B,1977ApJ...216..883B,1978ApJ...226.1087C},
or had self-consistently evolving open outer boundaries \citep{1991ApJ...370...60M};
in this latter case the initial conditions determined the velocity
dispersion and the density in the NSC outer region. Since no additional
boundary conditions or an outer thermal bath were introduced, the
conditions in the outer regions (where relaxation is slow) hardly
changed (see below), and the NSCs was effectively described as isolated
system, unrelated to the larger scale evolution of the galaxy. Thereby
it could not reproduce a configuration based on the ${\rm M-\sigma}$
relation today, unless arbitrarily put by hand as initial conditions.
Given these limitations and the debated understanding of MBH feedback
and the origin of the ${\rm M-\sigma}$ relation, we believe the ad-hoc
assumption we use is quite reasonable, and compared with other choices
it introduces a plausible relation between the NSC and the larger
scale environment consistent with current observations.\textbf{ }

Our modeling accounts for the NSC stellar population (the density
profile) up to the radius of influence in the evolving NSC, while
the stellar population outside the region is assumed to be consistent
with the large scale galactic bulge, as described above. We assume
the potential is dominated by the MBH and neglect the contribution
from the stellar population of the NSC, i.e. the contribution of the
stellar component up to the radius of influence is neglected, similar
to the original BW study. Up to the radius of influence this simplifying
assumption is justified, while far from the MBH it breaks down. Nevertheless,
at these regions the relaxation time becomes larger than the Hubble
time, and the background does not change due to the NSC internal evolution.
The intermediate regions beyond the radius of influence, in which
relaxation time can still be significantly shorter than a Hubble time
are small, and therefore our simplified modeling should still well
capture the overall dynamics of the NSC. 

For our final integrated TDH results (see Section \ref{sub:A-global-model}),
we extended the range of MBHs masses between $10^{5}$ and $5\times10^{7}$
${\rm M_{\odot}}$ by using the interpolated power-law function that
describes the dependence between TDE rates and MBH mass (Eq. \ref{eq:integ_fin_eq}).
We do not consider TDEs from lower mass MBHs, or TDEs of evolved stars,
which could be relevant even for MBHs above this mass range. 

Here we only consider simple models of spherical NSCs hosting non-evolving
MBHs of given masses with relaxation dominated by two-body scattering
of low mass (Sun-like) stars. Other models are beyond the scope of
this initial study and will be explored elsewhere.

\section{RESULTS \& DISCUSSIONS}

\subsection{Tidal disruption events history from Fokker-Planck analysis}

\subsubsection{Cluster infall and star-formation in NSC outskirts\label{sub:Cluster-infall-and SF}}

Stellar clusters infalling into the nuclear regions are typically
disrupted and shed their stars at scales of a few pc from the Galactic
center, due to their disruption by the potential of the MBH and the
existing NSC (e.g. \citealp{2014ApJ...784L..44P}). There is also
evidence for star-formation, young stellar populations and dense star-forming
clumps in these regions (e.g. the nuclear stellar disk in the Galactic
center, and young stellar populations in many galactic nuclei, \citealp{2006AJ....132.2539S,2011ApJ...732..120O}).
Together, these suggest that both in-situ SF and cluster infall scenarios
give rise to a source term of stars in regions at a few pc from the
MBH. In the following we present the results of FP calculations, where
star-formation/cluster infall scenarios are assumed to introduce new
stars, a ``source term'' in the FP models, in the regions between
2-3.5 pc. Note that we use the term SF throughout the following discussion,
but it effectively also represents a source term of stars from infalling
clusters.

\begin{figure*}[T!]
\includegraphics[scale=0.62]{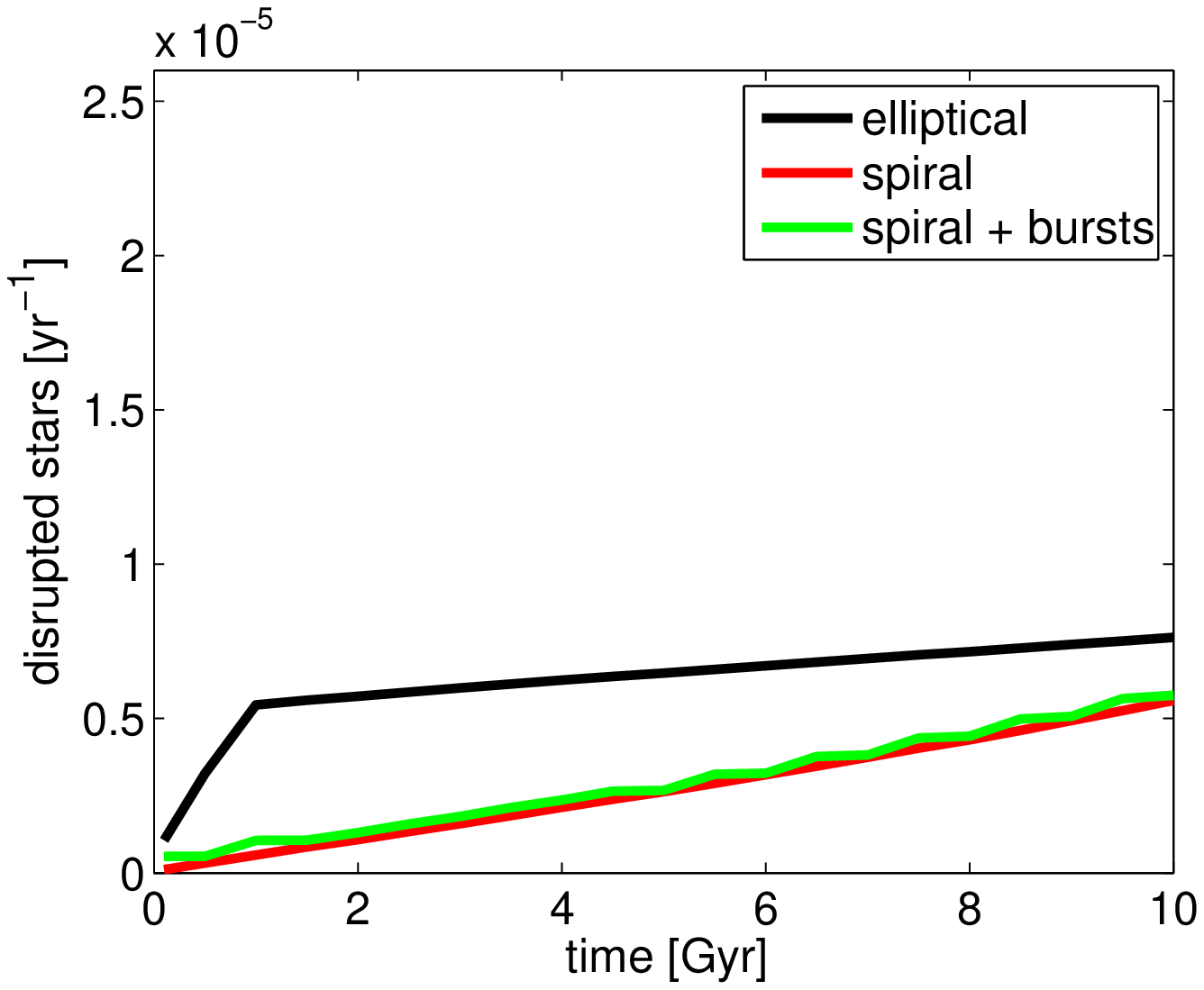}\includegraphics[scale=0.62]{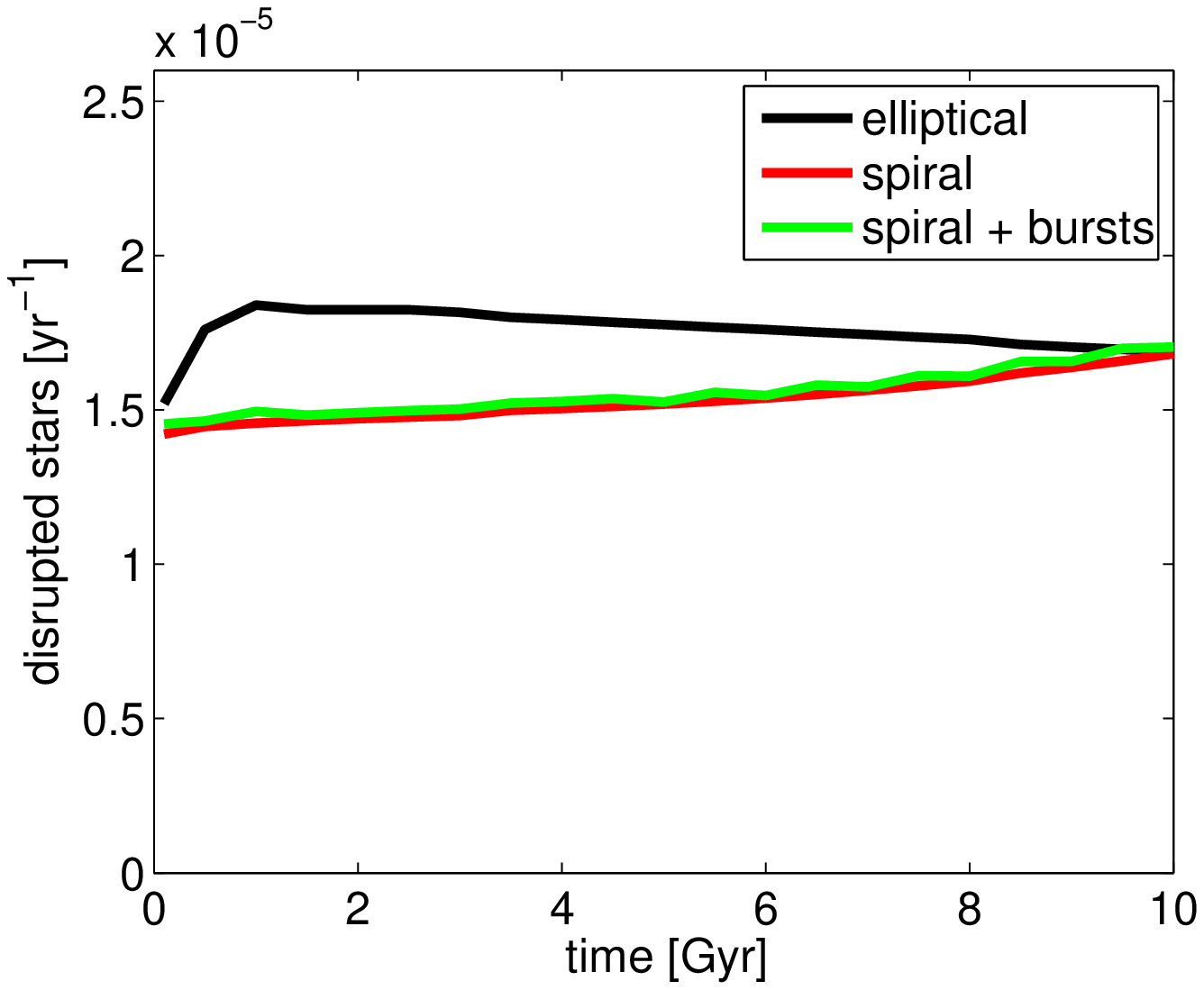}

\caption{The evolution of TDEs for NSCs with an MBH of $4\times10^{6}M_{\odot}$
for 3 types of galaxies, that are evolved through in-situ SF (left)
and from a primordial cusp (right). The rates of SF are $10^{-3}$
and $10^{-4}$ stars$\times$year$^{-1}$ for elliptical and spiral
galaxies respectively. The total number of stars formed throughout
the evolution is the same; NSC with primordial cusps have a larger
number of stars as they have similar SF histories but they also include
the stellar populations of the primordial cusps. \label{fig:cusp_3_galaxies_low_rate}}
\end{figure*}

\begin{figure*}[!]
\includegraphics[scale=0.62]{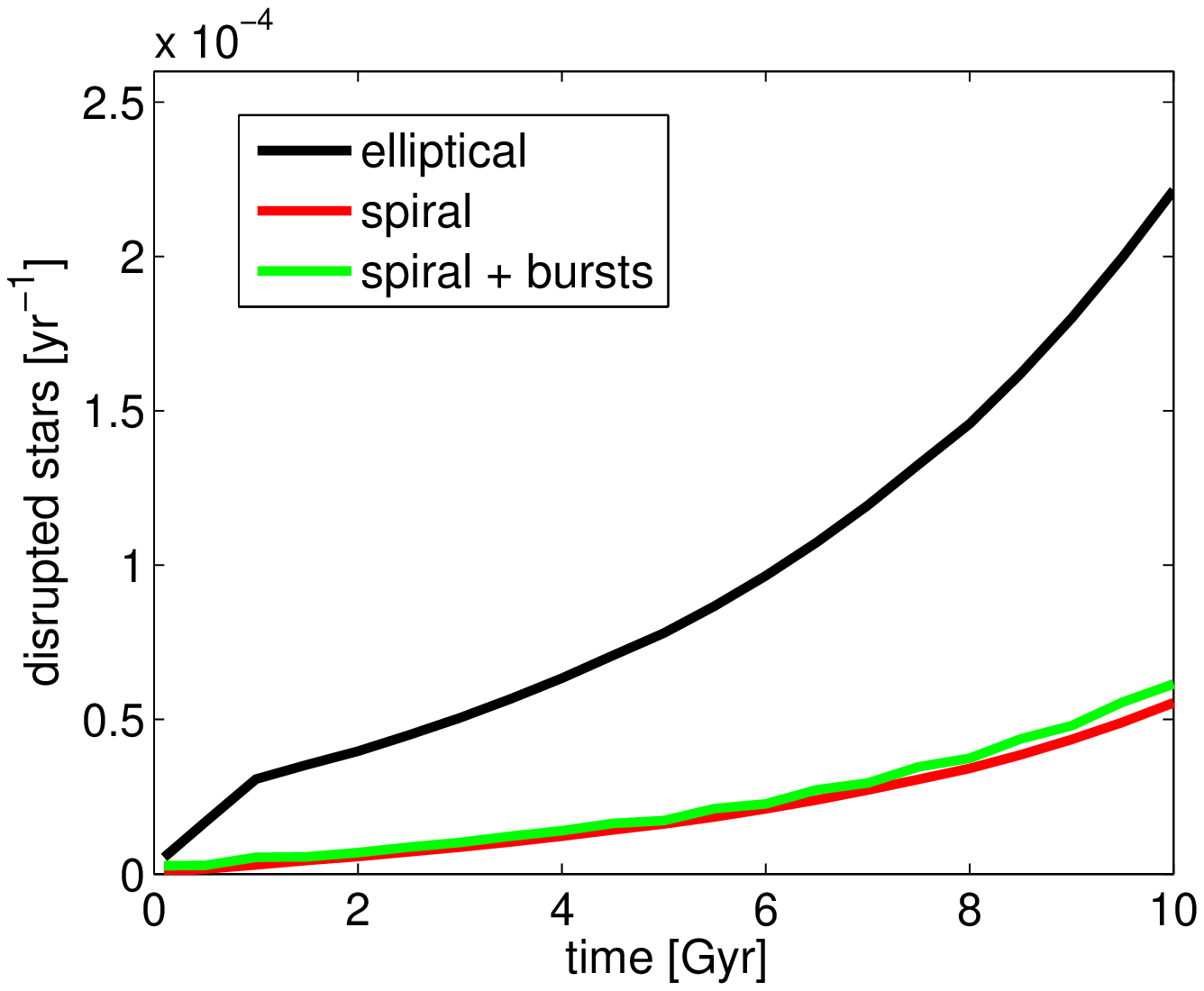}\includegraphics[scale=0.62]{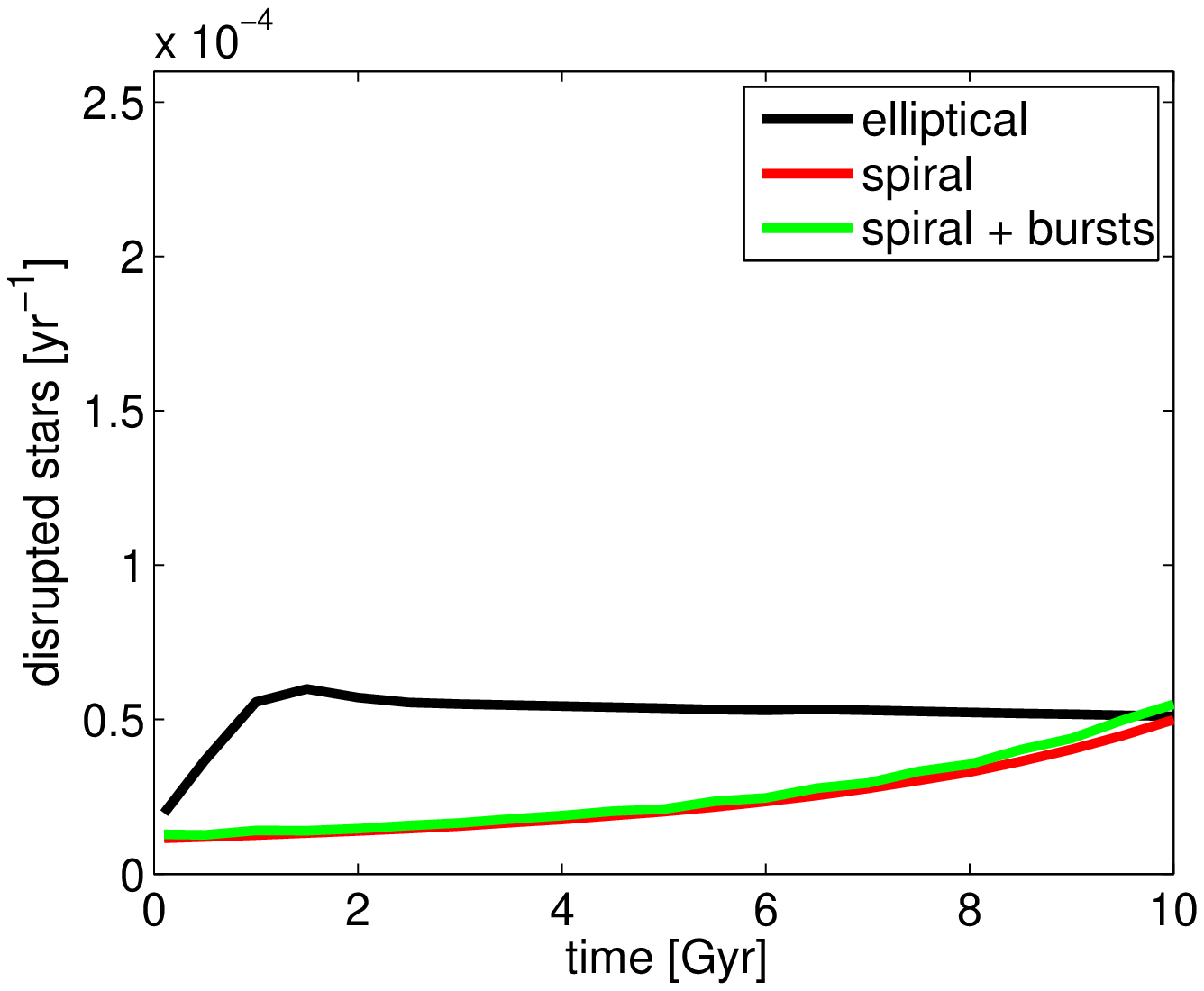}

\caption{The TDH as shown in Figure \ref{fig:cusp_3_galaxies_low_rate} for
higher rates of SF ($5\times10^{-3}$ and $5\times10^{-4}$ stars$\times$year$^{-1}$
for elliptical and spiral galaxies respectively). \label{fig:SF_3_galaxies_high_rate}}
\end{figure*}

As discussed in the previous sections, we have explored the TDE rates
as a function of the SF rate and characteristics. Figures \ref{fig:cusp_3_galaxies_low_rate}
and \ref{fig:SF_3_galaxies_high_rate} show the evolution of the TDE
rates with time for various models of the NSCs. As expected, the TDE
rates grow with time as the NSC stellar population and the density
of the NSCs increase due to star-formation. The early times TDE rates
in the models of early star-formation (ellipticals) are significantly
higher than the corresponding rates in the models with long term SF
(spirals). However, at later times the NSCs in both cases saturate
to similar levels as the NSCs reach similar densities at late times
(in cases of the same total cumulative SF throughout the evolution).
The TDE rates in NSCs with initial primordial cusps are generally
higher, since the overall stellar populations are larger. The rise
in the TDE rates is not linear as in the built-up NSCs models as relaxation
processes are more pronounced already at early stages. Since typical
relaxation times are longer than the length of SF bursts in our models,
it is not surprising that the SF bursts models behave very similarly
to those with continuous SF (with same averaged SF rates). Henceforth
we omit the results for SF burst models, which are essentially the
same as those in the continuous SF models. 

The high SF during the first Gyr in the elliptical galaxies models
give rise to a significant increase of the stellar population in the
NSC, and correspondingly leads to shorter relaxation times, and higher
densities. In turn the TDE rates rapidly increase during this epoch,
due to these effects, and the increase becomes shallower only at later
times after the end of the SF epoch. Nevertheless, the rates continue
to grow in time even after 10 Gyr of evolution as stars from the galaxy
slowly diffuse into the NSC. In the spiral galaxy models SF is continuous
throughout the evolution, and the TDE rates similarly increase in
a more continuous manner. 

\begin{figure*}[!]
\includegraphics[scale=0.62]{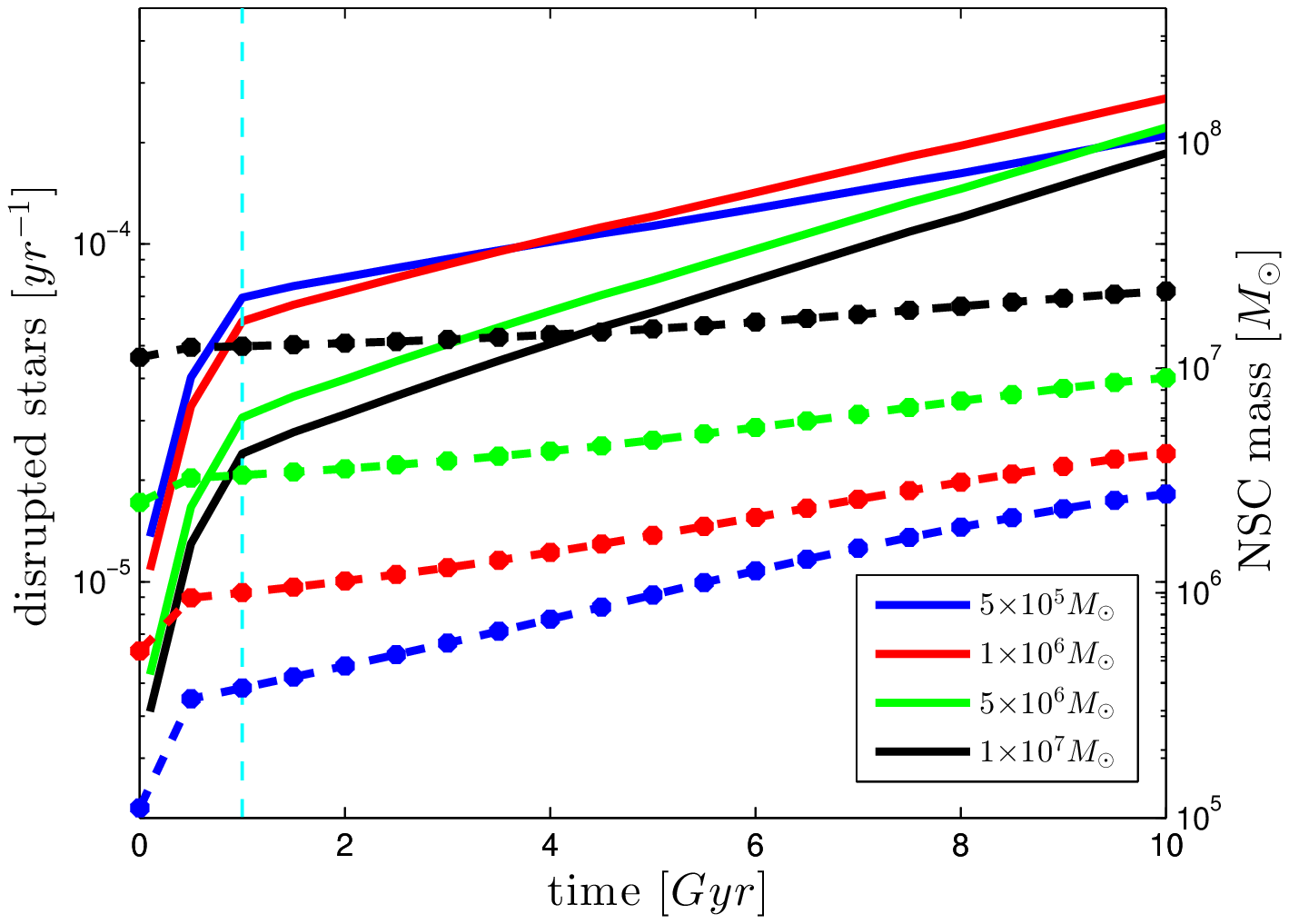}\includegraphics[scale=0.62]{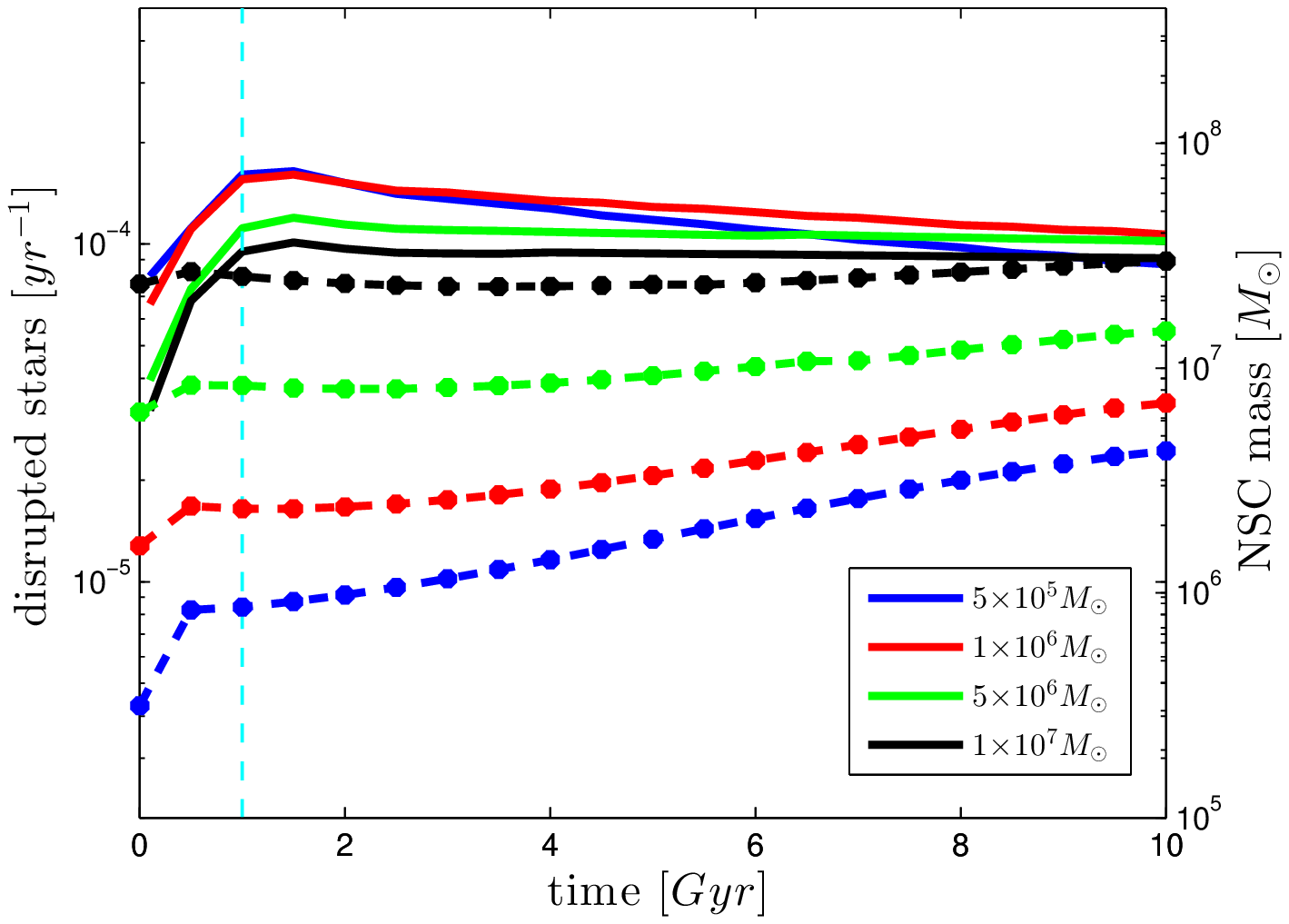}\caption{Evolution of TDE rates (continuous lines) for four different masses
of MBHs for model $E$ along with mass of the NSC (dashed lines).
The cyan vertical dashed line marks the epoch at which the SF stops.
NSC masses and the corresponding TDE rates of the same models are
marked with the same colors. Left: TDH for NSC evolved from built-up
star formation. Right: TDH for NSC evolved from a primordial cusp.\label{fig:MBH_TDE1}}
\end{figure*}

\begin{figure*}[!]
\includegraphics[scale=0.62]{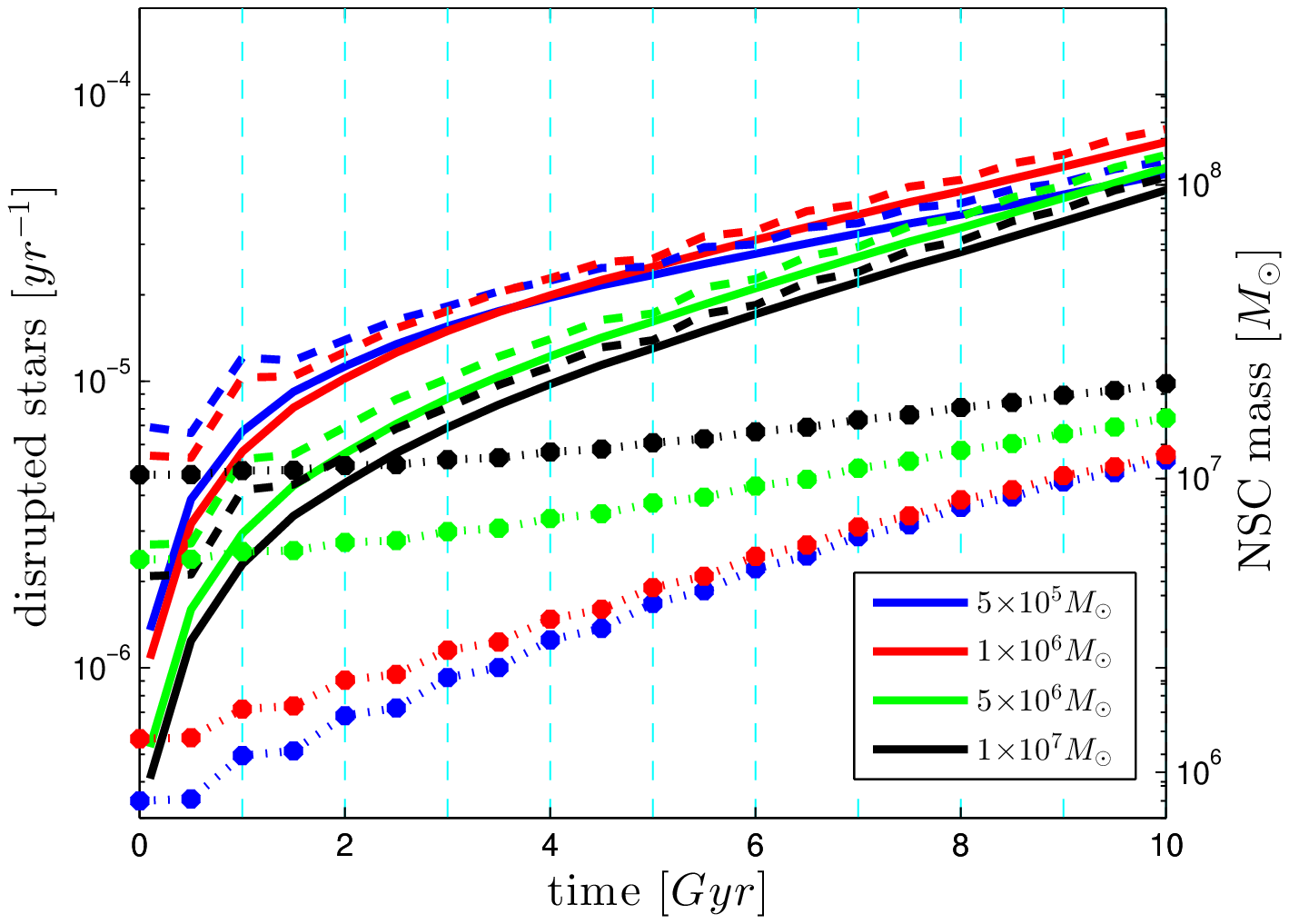}\includegraphics[scale=0.62]{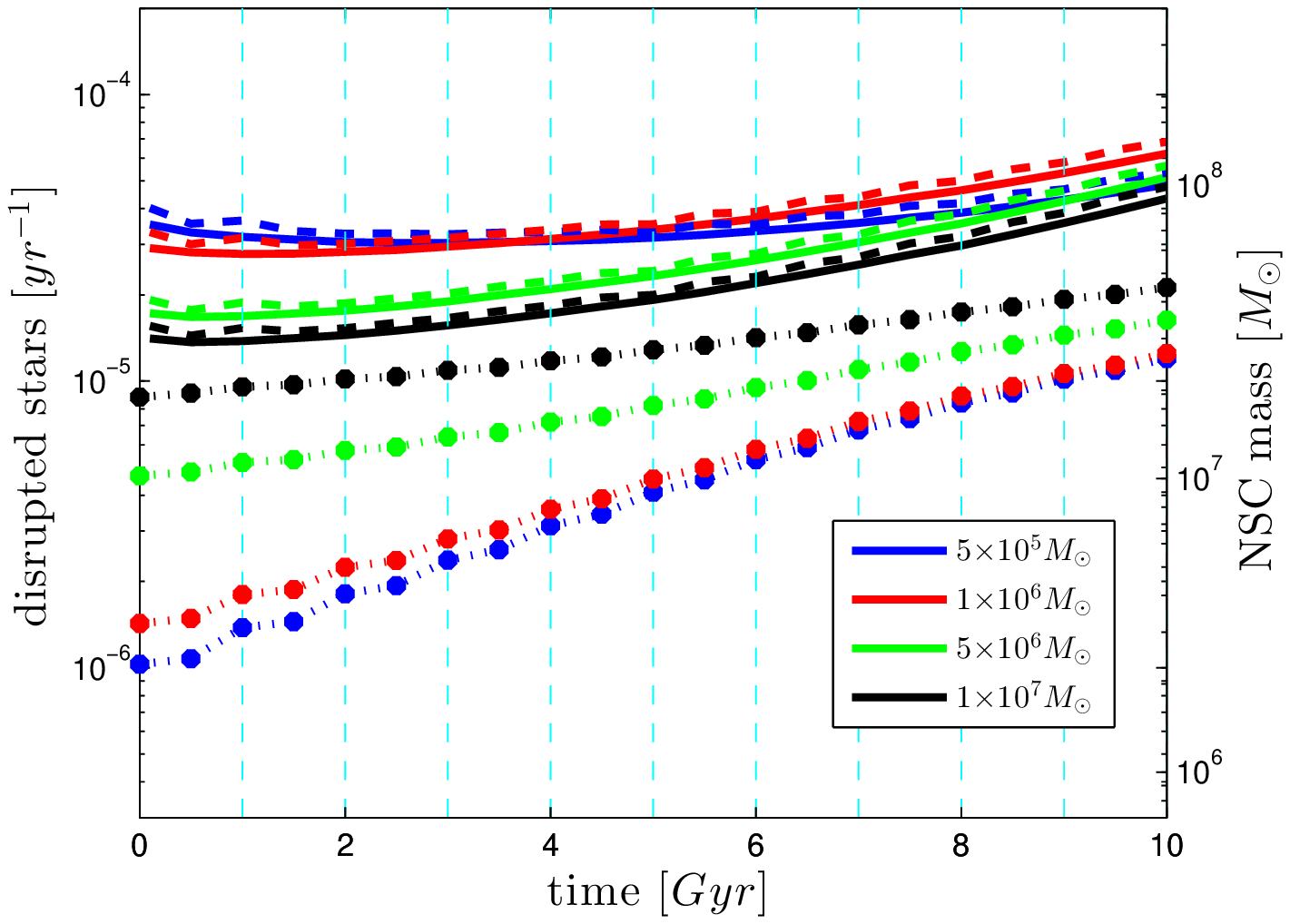}

\caption{Evolution of TDE (continuous lines) rates for four different masses
of MBHs for models $S_{0}$ (solid lines) and $S_{B}$ (dashed lines).
NSC masses and the corresponding TDE rates of the same models are
marked with the same colors. The blue dashed lines mark the time of
the SF bursts (every 1 Gyr). Left: TDH for NSC evolved from built-up
star formation. Right: TDH for NSC evolved from a primordial cusp.\label{fig:MBH_TDE2}}
\end{figure*}

In addition to the comparison between TDEs in different types of galaxies,
we compare the TDE rates for different masses of MBHs ($5\times10^{5}M_{\odot}$,
$1\times10^{6}M_{\odot}$, $5\times10^{6}M_{\odot}$ and $1\times10^{7}M_{\odot}$).
We present the behavior of each scenario and for each of the MBH masses
along with the evolution of the total mass of the NSC; see Figures
\ref{fig:MBH_TDE1} and \ref{fig:MBH_TDE2}. These probe the dependence
of the TDE rates on the mass of the MBHs. It can be clearly seen that
the TDE rates decrease with the mass of the MBH. This result has already
been discussed by \citet{Wang2004}, and can be well understood from
loss cone analysis (see \citealt{2013CQGra..30x4005M}). 

\citet{Merritt2009} have discussed the evolution of TDE rates with
time in the case of a non-SF NSC. Though our models always include
SF, the evolution of models of pre-existing NSCs in which only an
early SF epoch occurs, evolve very similarly to the models explored
in \citet{Merritt2009}, after the end of the SF. Such models generally
predict a global decrease of TDE rates over time. . A different trend
exists in the evolution of TDE rates in models with long-term SF,
in which the TDE rates slowly increase with time as the NSC keeps
growing due to SF. 

\begin{figure*}[!]
\includegraphics[scale=0.62]{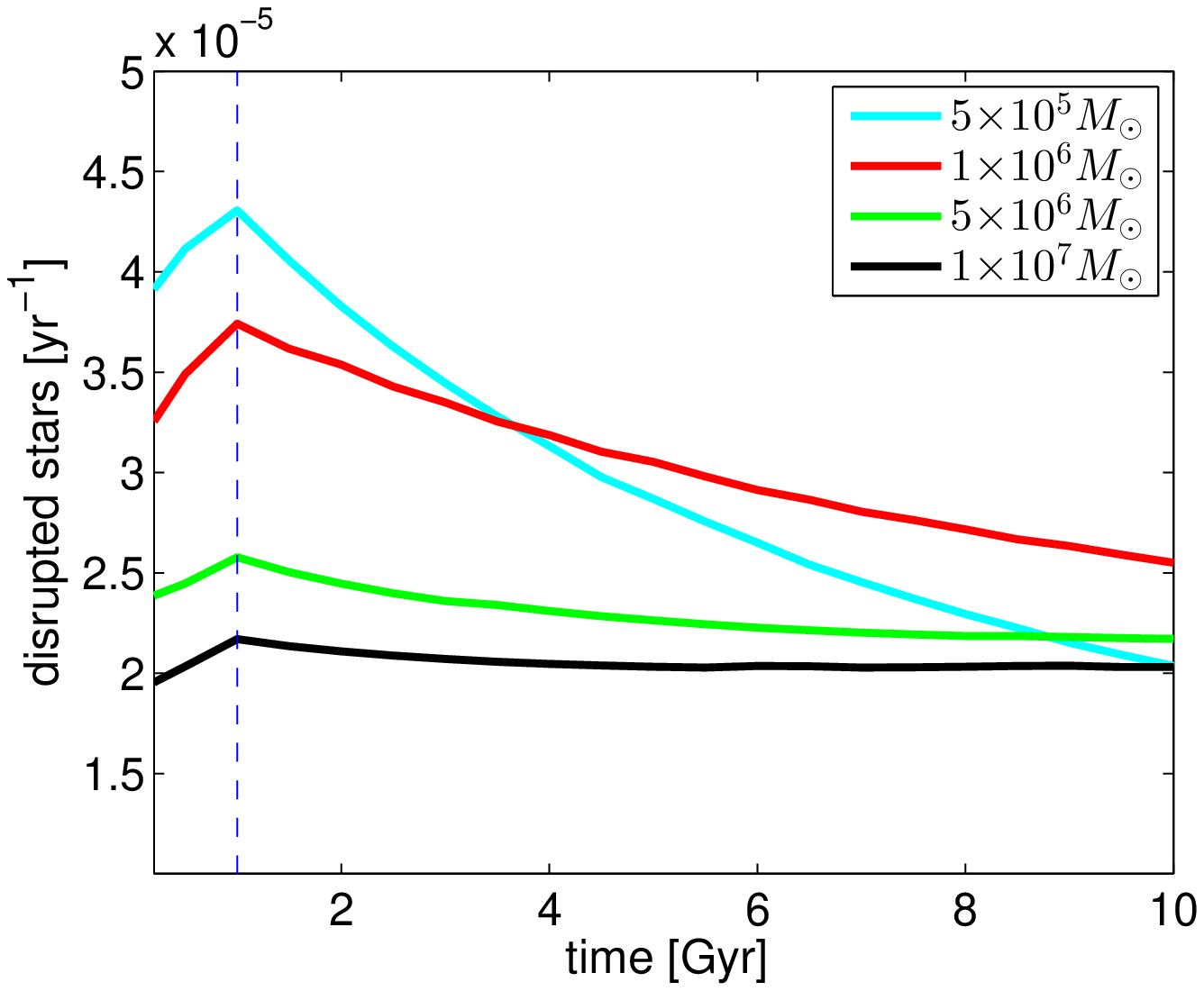}\includegraphics[scale=0.62]{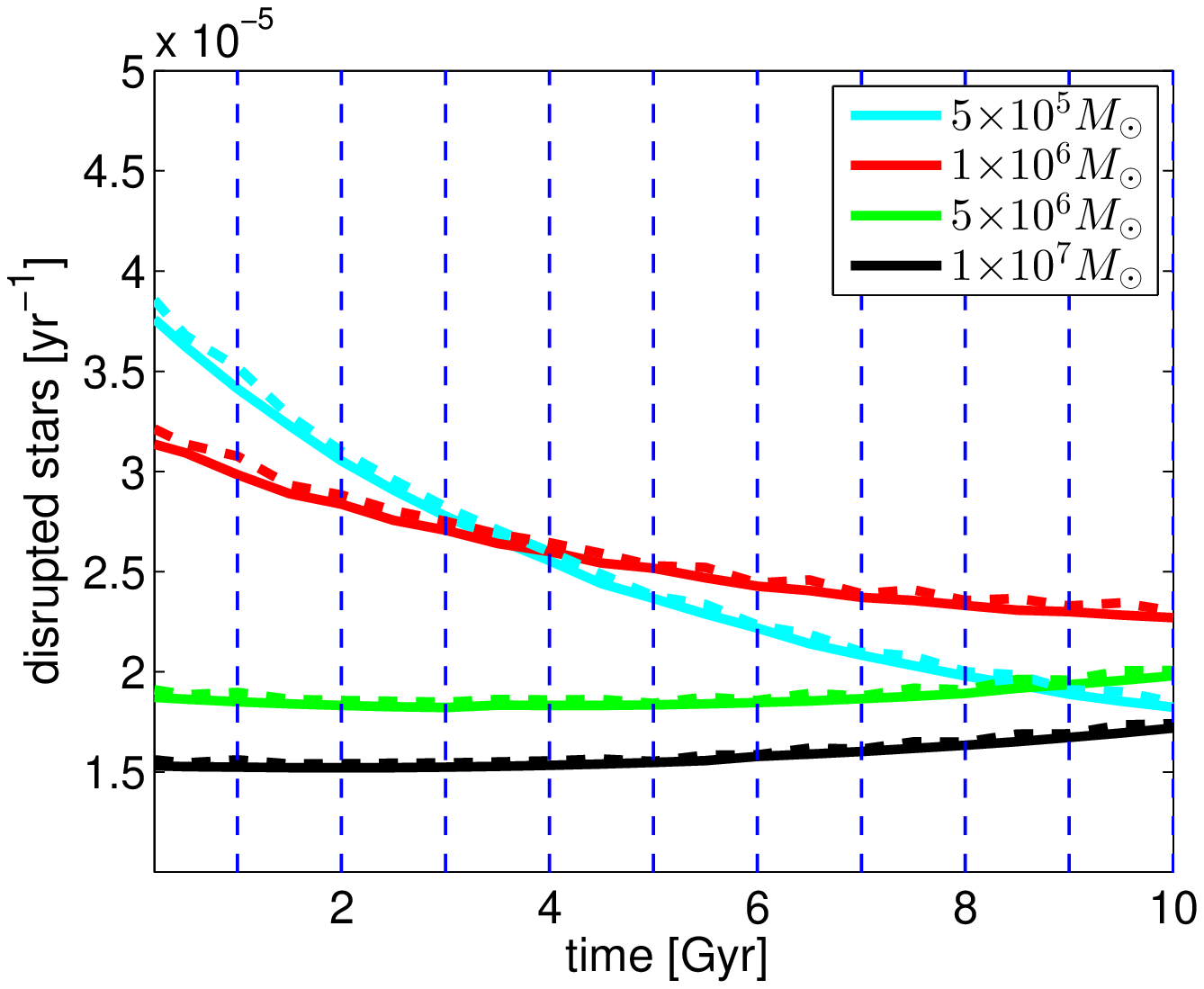}

\caption{Evolution of TDE (continuous lines) rates for four different masses
of MBHs for models $E$ (left panel) and $S_{0}$ $S_{B}$ (right
panel) for lower rates of SF. The dashed blue lines mark the epoch
at which SF stops and the bursts of SF for $E$ and $S$ models respectively.\label{fig:Evolution-of-TDE_lower-rates-SF}}
\end{figure*}

Fig. \ref{fig:Evolution-of-TDE_lower-rates-SF} shows the TDE rates
for the models with lower SF rates and low MBH masses. Note that in
the models with MBHs masses of $5\times10^{5}M_{\odot}$ and $10^{6}M_{\odot}$
the TDE rates decay with time. In these cases the restive increase
in the MBH mass significantly increase the relaxation time leading
to lower TDE rates, even though the total masses of the NSCs increase. 

\subsubsection{Star-formation in the inner regions of NSCs}

\begin{figure*}[!]
\includegraphics[scale=0.62]{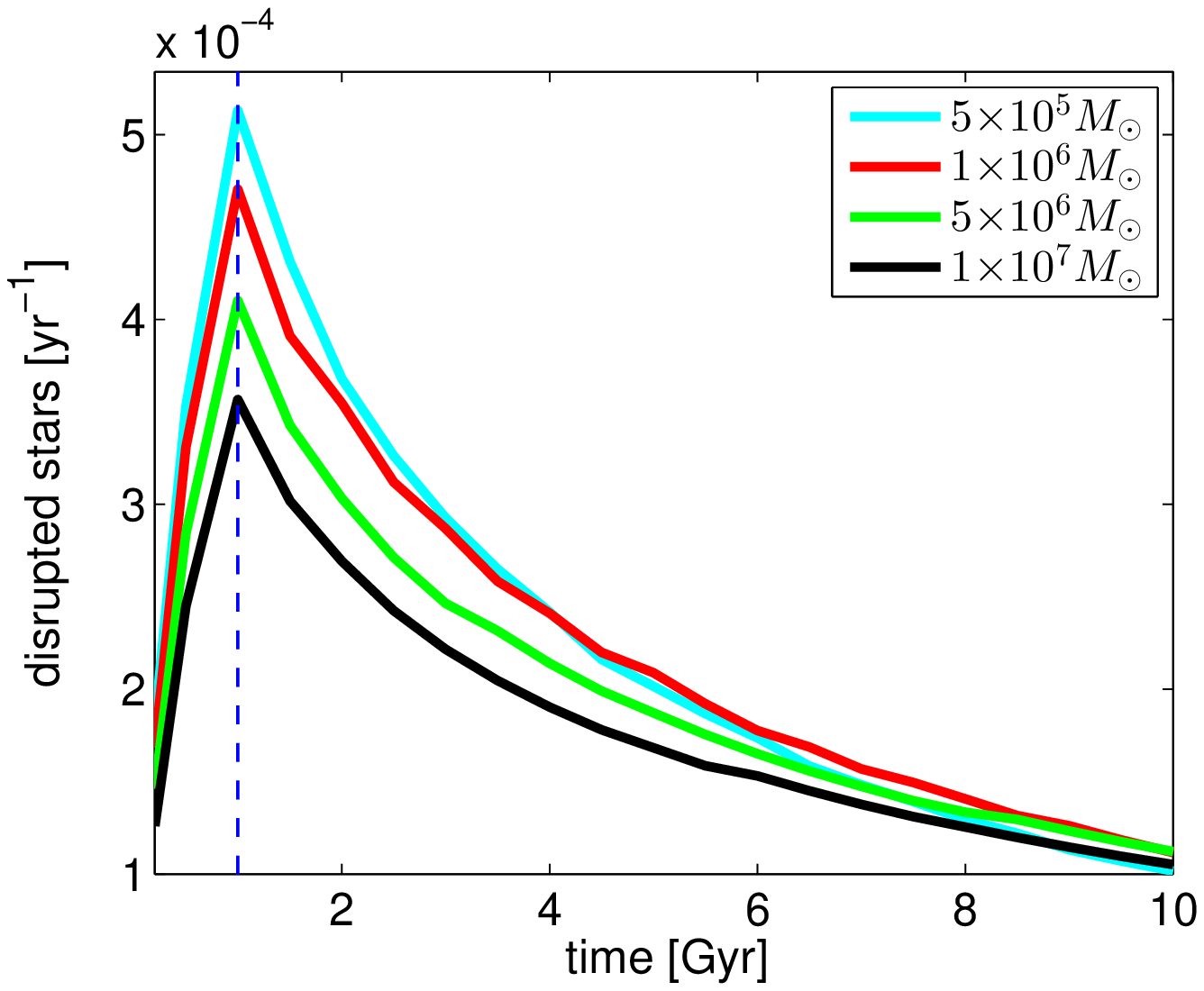}\includegraphics[scale=0.62]{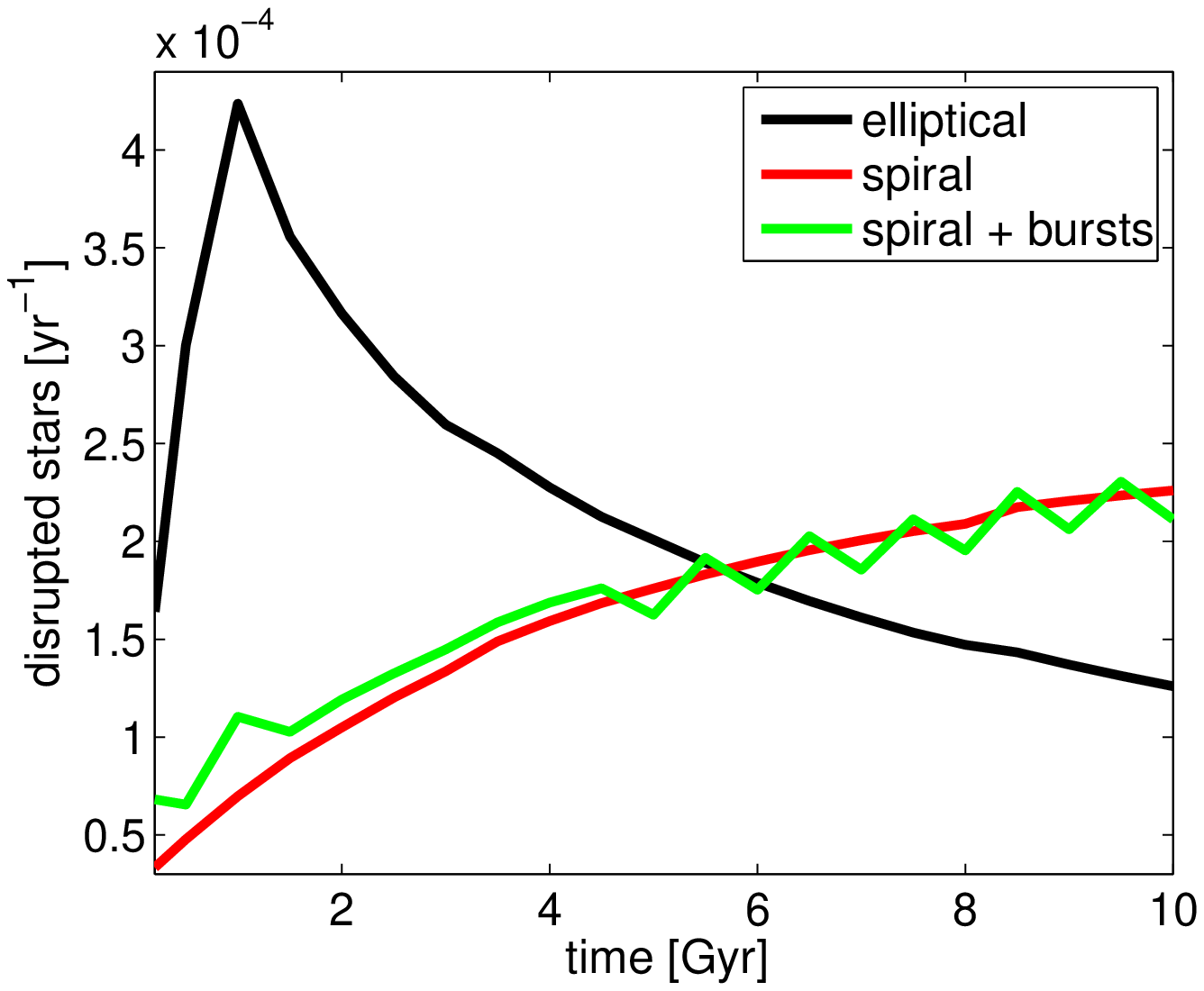}

\caption{TDEs history for scenarios in which SF that takes place in the inner
parsec (0.05-0.1pc), for NSCs that are built up through in-situ SF.
Left: TDH evaluated for elliptical galaxies (SF occurs only in the
first Gyr) for four masses of MBH. Right: TDE histories for three
different types of galaxies for high SF rates ($5\times10^{-3}M_{\odot}$,
similar to Figure \ref{fig:SF_3_galaxies_high_rate}) \label{fig:TDEs-Diff_Rng}}
\end{figure*}

In addition to the TDH presented above, corresponding to SF at a distance
of $2-3.5{\rm pc}$, we also studied the TDH in a different scenario,
where SF is assumed to mainly occur in the inner regions of NSCs.
Evidence for such more nucleated SF can be found in our own galaxy,
where a young stellar population is observed in the inner $\sim0.1$pc
near the MBH; though it is not clear if such SF is typical (such regions
can not yet be resolved in other galaxies), we consider this possibility
for completeness. We\textbf{ }followed the same conditions used in
the previous scenarios, but we changed the range of SF to correspond
to distances of $0.05-0.1{\rm pc}$. Figure \ref{fig:TDEs-Diff_Rng}
presents the TDEs history for SF rate of $5\times10^{-3}yr^{-1}$
and $5\times10^{-4}yr^{-1}$ for elliptical and spiral galaxies, respectively.
On the right panel we notice a different behavior\textbf{ }of the
TDH for elliptical galaxy compared with the outskirt SF scenarios:
while there is a rapid growth in the disruption rates during the epoch
of SF, the rates decreases with time once the\textbf{ }SF process
is quenched. Since the stellar densities increase on relatively short
times as many stars are formed close to the MBH, the NSC is initially
far from its steady state configuration. The inner regions are ``over-populated''
and relaxation processes take time to redistribute these stars throughout
the NSC. Since stars in these regions are more susceptible to being
tidally disrupted (larger loss cone), the TDE rates increase rapidly,
before the NSC structure re-equilibrates. The long-term SF occurring
in spiral galaxies allows for relaxation processes to redistribute
the stars, and the TDE rates increase more slowly, as the NSC is built-up.

\subsubsection{Global SF in NSCs}

\begin{figure*}[!]
\includegraphics[scale=0.62]{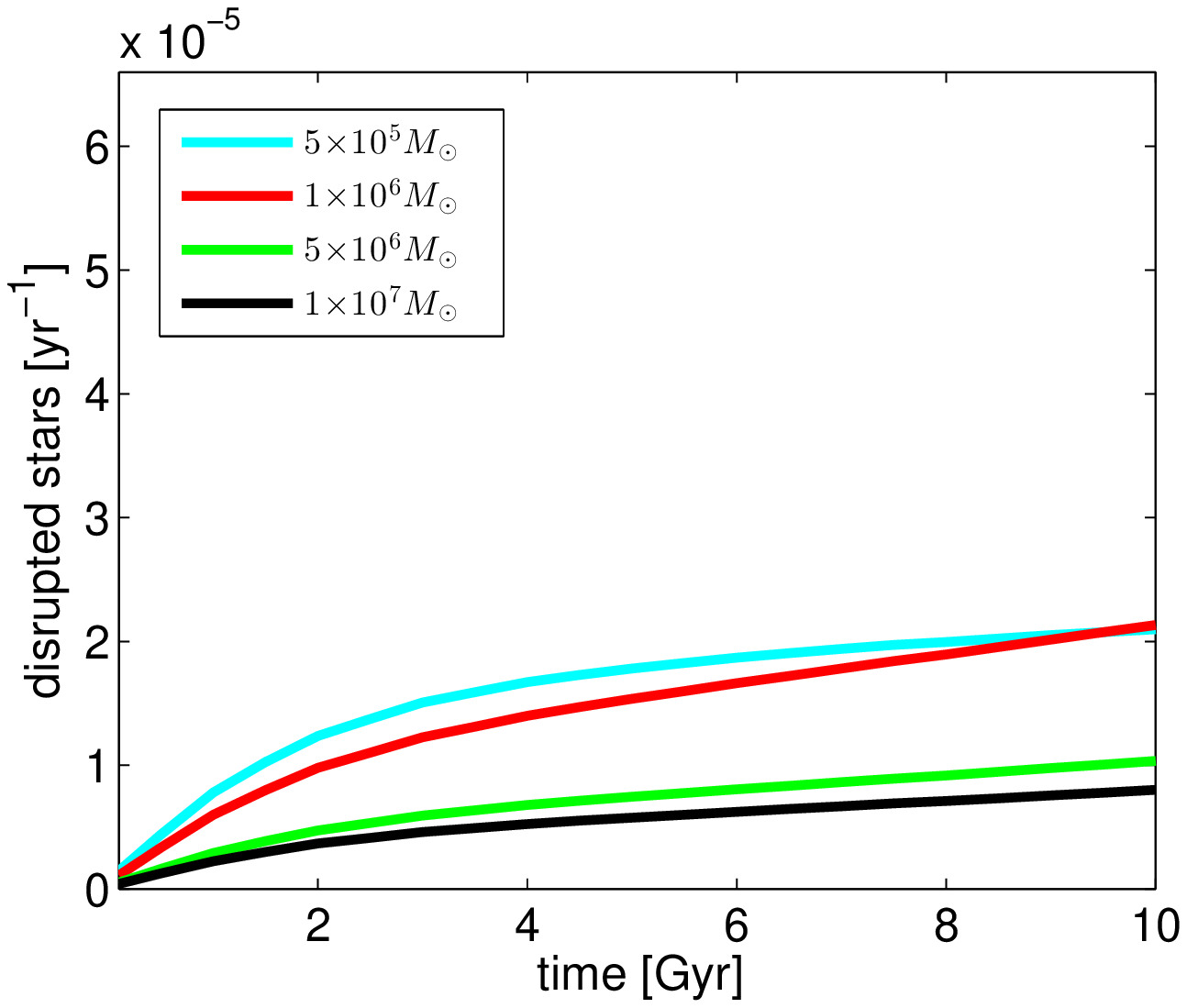}\includegraphics[scale=0.62]{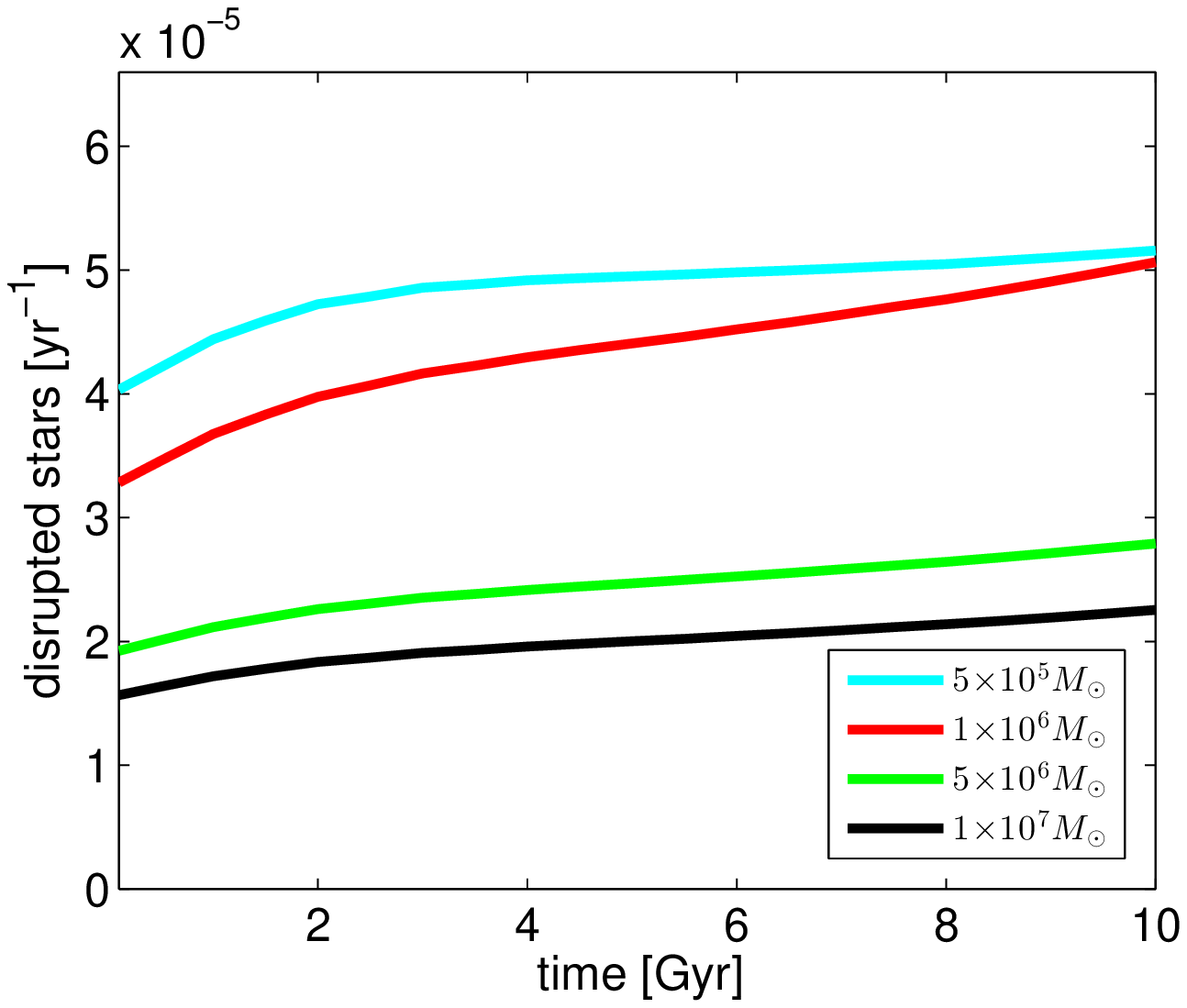}\caption{Evolution of TDE rates for four models with different masses of MBHs,
and a SF history which follows the global universal SF models \citep{2014ARA&A..52..415M}
normalized to the NSCs. (the total number of added stars to each model
is equal to the total number of stars in models shown in Fig. \ref{fig:cusp_3_galaxies_low_rate}.
Left: TDH for NSC evolved from built-up global star formation. Right:
TDH for NSC evolved from a primordial cusp with global SF. \label{fig:Evolution-of-TDE-global}}
\end{figure*}

In addition to the TDHs of NSCs evolving through an early SF epoch
or through continuous SF (described in Table \ref{tab:galaxies_prop}),
we also studied models in which the SF follows the global universal
SF history (i.e. \citealt{2014ARA&A..52..415M}) in such global models
early SF is more significant, and then it gradually decrease by an
order of of magnitude (\citealt{2014ARA&A..52..415M}). Such models
are therefore expected to be more similar to the early SF epoch models
(``elliptical galaxies''). In Fig. \ref{fig:Evolution-of-TDE-global}
we show the results from thee models, which confirm these expectations;
these models show similar trends as those in Fig. \ref{fig:Evolution-of-TDE-global}. 

\subsection{Tidal disruption events history using N-body simulation data}

Following the cluster infall study (described in \citealt{2014ApJ...784L..44P}),
we evaluated the disruption rates of stars based on data obtained
from N-body simulations. This provides us with a complimentary approach
to the FP calculations, through a different type of analysis. As we
show in the following the FP calculation for models with SF in the
NSC outskirts and the calculations based on N-body data are generally
comparable. Note that the N-body calculations are computationally
expensive, and therefore we show only a single case, while our general
main results (shown in the previous sections) are based on the FP
calculations. Note that direct calculation of the TDE rates from an
N-body simulation requires appropriate resolution, number of particles
and calibration. We have followed a more simplified approach where
the N-body simulations only provide us with the effective radial density
profile at any given point in time, from which we then calculate the
TDE rate semi-analytically, similar to the approach used in Mastrobuono-Battisti,
Perets \& Loeb (2014). This avoids the obstacles arising from the
direct calculations of TDE rates in N-body simulations, at the cost
of a less accurate result, and the use of the simplifying assumption
of an isotropic distribution.

\begin{figure}[!]
\includegraphics[scale=0.62]{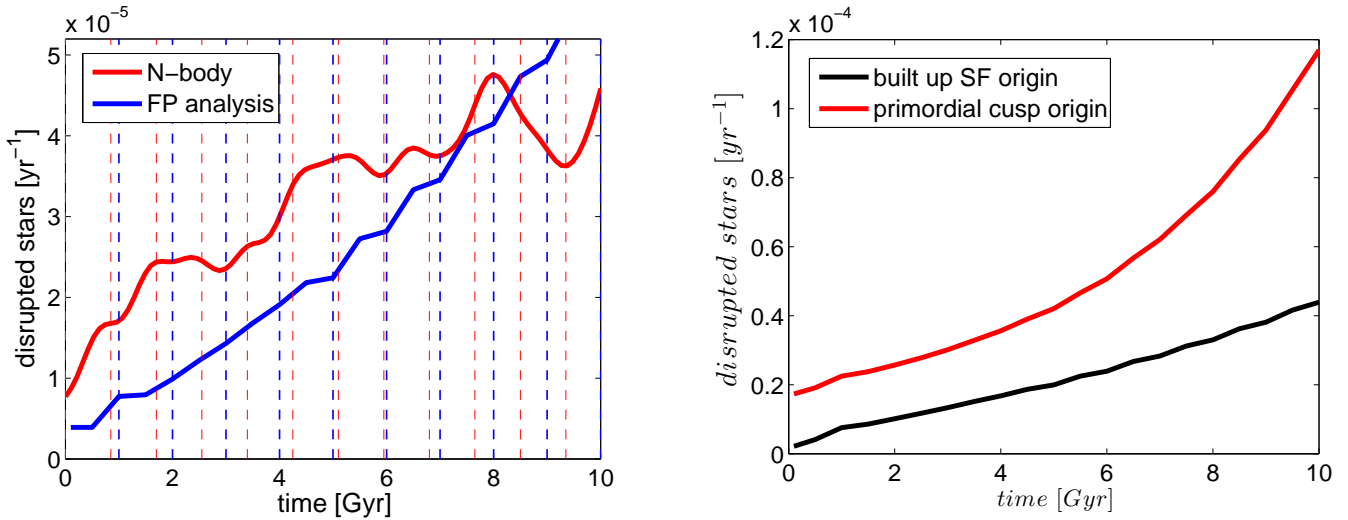}\caption{Comparison between TDEs history obtained from N-body simulations and
FP analysis. TDH for an NSC formed through clusters infall, compared
with the TDE history for the FP models with ``SF'' bursts. The red
dashed lines mark the time of the infall (N-body) and the blue dashed
lines mark the time of the SF bursts. \label{fig:N-body_RAND_FP}}
\end{figure}

In the cluster-infall scenario clusters are added to the system on
equally spaced time intervals. They then inspiral and disrupt, and
their stars build-up the NSC. To a large extent this is captured by
the FP models which introduce stars in the NSC outskirts; since the
infalling clusters disperse in the same distance scales (few pcs),
they introduce stars in the same manner as assumed in these FP models.
We therefore compare the result of these two scenarios in Figure \ref{fig:N-body_RAND_FP}.
As can be seen, both scenarios provide very similar results, suggesting
that the FP calculations can also be used to model the cluster-infall
scenario. 

\subsection{A global model for TDE history\label{sub:A-global-model}}

As described in the previous sections, we have followed the evolution
of the disruption rates of stars for different types of galaxies and
evolutionary/SF-history scenarios. Combining the rates we obtained
for each galaxy type with the observational information about the
fraction of each morphological galaxy type can provide a general prediction
for the TDH of the universe. This analysis is done as follows; for
each type of hosted MBH mass we obtain the global TDH by integrating
the rates we obtained for the different galaxies and MBH mass, weighted
by their relative fraction

\begin{equation}
\Gamma_{G}=\int_{M_{min}}^{M_{max}}\sum_{i}\Gamma_{G_{i}}(m_{BH})n_{G_{i}}(m_{BH})dM_{BH}\label{eq:integ_fin_eq}
\end{equation}
where $\Gamma_{G_{i}}$ is the TDE rate for the $i$-th galaxy type
with an MBH of mass $m_{BH}$, and $n_{G_{i}}(m_{BH})$ is its fraction
of such galaxies in the local universe. The data for the fraction
of each galaxy type is taken from observations (\citealt{Calvi2012})
and the MBH mass function is based on \citet{2012AdAst2012E...7K}.
Note that in practice we use a sum over the MBH masses for which we
have made the full calculation, rather than a continuous mass function.
We generally find that the TDE rate dependence on the MBH mass goes
approximately as $\Gamma_{TDE}^{avg}=c_{1}M_{\bullet}^{-\gamma_{1}}$,
where $\gamma_{1}=0.44$. We separate the rate dependence results
for each galaxy type, and find that the MBH mass goes as $\Gamma_{TDE}^{E}\sim M_{\bullet}^{-0.31}$,
and $\Gamma_{TDE}^{S}\sim M_{\bullet}^{-0.62}$, for elliptical and
spiral galaxies respectively. These findings of lower TDE rates obtained
for higher mass MBHs are not surprising and have already been seen
and discussed in depth in other studies. Previous models had not considered
continuous SF, however our models for early burst of SF (elliptical
galaxies) with no later SF are the most comparable. Indeed in those
cases we find consistent results; $\Gamma_{TDE}^{E}\sim M_{\bullet}^{-0.31}$;
where the range of the power law dependence found in previous studies
is $0.2\apprle\gamma_{1}\apprle0.5$ (e.g. \citealp{Wang2004,2013degn.book.....M,Vasiliev2013,2016MNRAS.455..859S}).

\begin{figure}[!]
\includegraphics[scale=0.62]{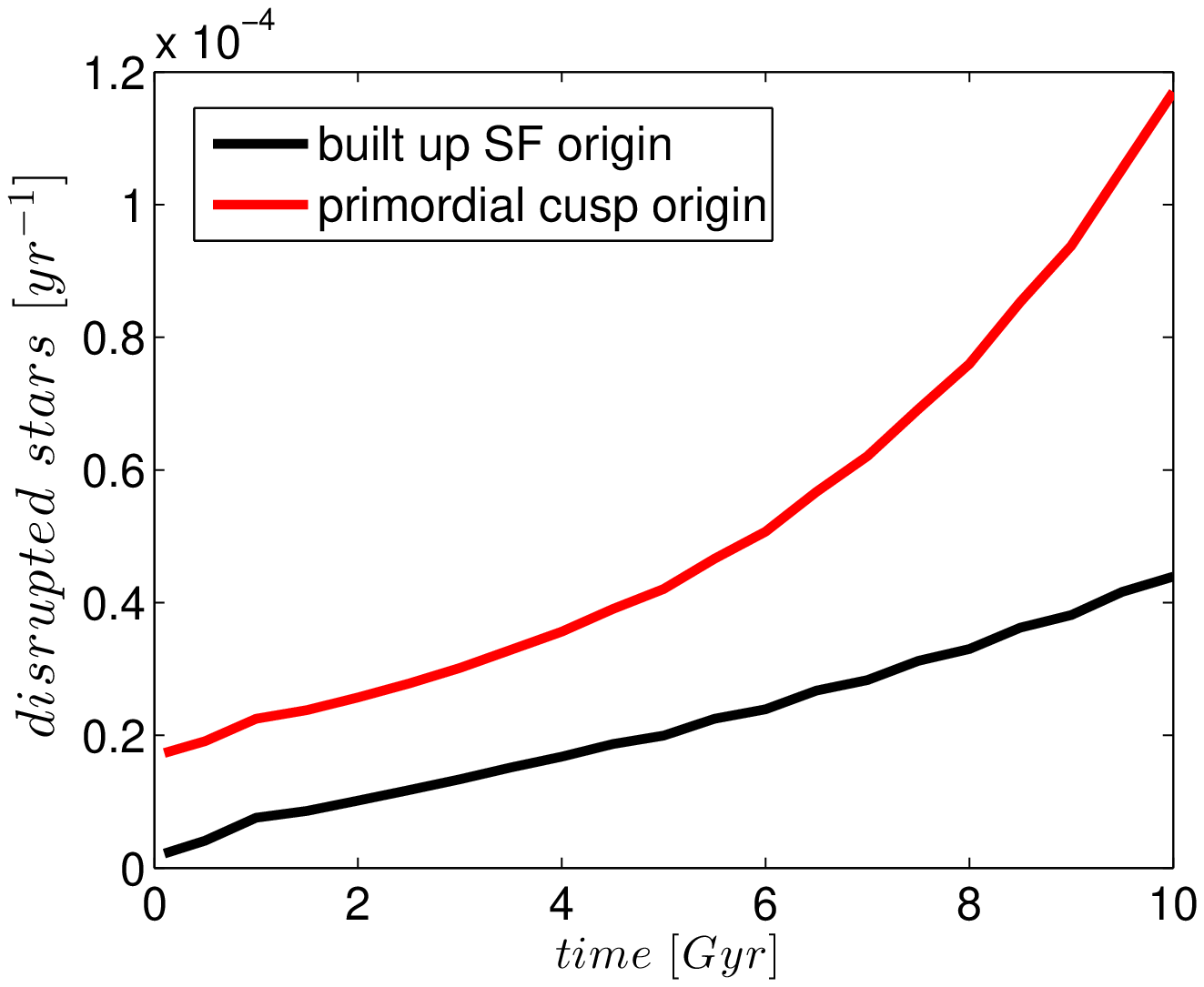}

\caption{The integrated cosmological history of TDEs rates per galaxy for the
built-up and the primordial cusp scenarios for NSC formation and evolution.
Note that for these final results we integrated the TDEs rates over
MBHs masses between $10^{5}M_{\odot}$ and $5\times10^{7}M_{\odot}$.
\label{fig:integrated_mass_total}}
\end{figure}

\section{SUMMARY}

In our work we provide a first theoretical study of the TDE rate history
which considers star-formation and cluster infall scenarios as models
for nuclear stellar cluster formation and evolution. We find that
in most scenarios the TDE rates grow with time, due to the growth
of the NSCs stellar density during their evolution, where the TDE
rates decrease inversely with the mass of the MBH they host. We explored
two different origins of the NSC: pre-existing NSC cusp that forms
very early in the galaxy evolution or an NSC that grows more slowly
through star formation or cluster infalls. NSCs with primordial cusps
have initially larger stellar populations. The overall TDE rates they
produce are therefore higher than those calculated for the cases of
NSC which are built-up on longer timescales, and they are non-negligible
already at early times. Consideration of different types of SF histories,
corresponding to elliptical and spiral galaxies, also show a similar
trend: the earlier the SF occurs (e.g. in elliptical where most SF
is assumed to occur at the first Gyr), the faster the NSC is built,
and the faster the TDE rates increase.

We also compared our disruption rates for NSCs with SF which occurs
in the inner parsec, and found that in such SF scenarios a the TDE
histories show a different behavior where: during the SF epoch the
TDE rates rapidly grow, but then decrease and decay to a much lower
level on a timescale of a few Gyrs.

Finally, to complement our main results from the Fokker-Planck models,
we also modeled the TDH using data from full N-body simulations of
the cluster-infall scenario. We found a good match between these rates
and the FP results corresponding to a source term of stars in the
NSC outskirts, further confirming the more simplified FP analysis. 

Our results are mainly based on simple 1D Fokker-Planck models. Although
these appear to be able to well capture the results of N-body simulations
(at least for the cluster-infall scenario modeled through N-body simulations),
they did not include a detailed galaxy merger history, and more complex
non-spherical structures of NSCs. The role of these aspects and their
effects on the TDH are not well determined, and should be considered
in future work

Future surveys of TDEs (e.g. using the LSST data) would be able to
provide not only better estimate of the TDE rates in the local universe,
and their dependence on the host galaxy, but also the history of TDEs.
As we have shown here, the TDH can provide us with information on
the overall evolution of nuclear stellar clusters and the star-formation
history and properties in galactic nuclei through the combination
of the observational data and the theoretical studies.

\acknowledgements{}

We would like to thank Clovis Hopman for the use of the basic components
in his FP code for developing the FP code used in our simulations.
We would also like to especially thank the anonymous referee
for important comments and suggestions that considerably improved
the manuscript. We acknowledge support from the Technion Asher Space
Research Institute and the I-CORE Program of the Planning and Budgeting
Committee and The Israel Science Foundation grant 1829/12. 

\bibliographystyle{plainnat}

\end{document}